\documentclass[12pt]{article}
\usepackage{amsmath,amsfonts,amssymb,dcolumn,lscape,subfigure,alltt,authblk}
\usepackage{graphicx,color,soul}
\usepackage{tabularx}
\usepackage[utf8]{inputenc}
\usepackage[T1]{fontenc}

\begin{document}
\baselineskip=6.5mm
\bibliographystyle{aip}
\newcommand{\be}{\begin{equation}}
\newcommand{\ee}{\end{equation}}
\newcommand{\Be}{\begin{eqnarray}}
\newcommand{\Ee}{\end{eqnarray}}
\def\lg{\langle}
\def\rg{\rangle}

\def\b{\beta}

\def\cal{calix[4]arene }
\def\cals{calix[4]arenes }
\def\bAla{$\b$-alanine octamer}

\parindent 0cm

\author[]{Marco Oestereich}
\author[]{J\"urgen Gauss}
\author[]{Gregor Diezemann}
\affil[]{Department Chemie, Johannes Gutenberg-Universit\"at Mainz, Duesbergweg 10-14, 55128 Mainz, Germany}
\title{Adaptive Resolution Force Probe Simulations: Coarse Graining in the Ideal Gas Approximation}
\maketitle
\begin{abstract}
The unfolding of molecular complexes or biomolecules under the influence of external mechanical forces can routinely be simulated 
with atomistic resolution.
To obtain a match of the characteristic time scales with those of experimental force spectroscopy, often coarse graining procedures are employed.
Here, building on a previous study, we apply the adaptice resolution scheme (AdResS) to force probe molecular dynamics (FPMD) simulations using two model systems as examples.
One system is the previously investigated calix[4]arene dimer that shows reversible one-step unfolding and the other example is provided by a small peptide, a $\beta$-alanine octamer in methanol solvent.
The mechanical unfolding of this peptide proceeds via a metastable intermediate and therefore represents a first step towards a  complex unfolding pathway.
In addition to increasing the complexity of the relevant conformational changes we study the impact of the methodology used for coarse graining.
Apart from a standard technique, the iterative Boltzmann inversion method, we apply an ideal gas approximation and therefore we replace the solvent by a non-interacting system of spherical particles.
In all cases we find excellent agreement between the results of FPMD simulations performed fully atomistically and the AdResS simulations also in the case of fast pulling.
This holds for all details of the unfolding pathways like the distributions of the characteristic forces and also the sequence of hydrogen-bond opening in case of the $\beta$-alanine octamer.
Therefore, the methodology is very well suited to simulate the mechanical unfolding of systems of experimental relevance also in the presence of protic solvents.
\end{abstract}
%
%
\newpage
\section*{I. Introduction}
The application of mechanical forces to study the unfolding of biomolecules or molecular complexes like protein-ligand systems has become a standard way to study the details of the kinetics of conformational changes\cite{Janshoff:2000,Kumar:2010,Zoldak:2013,Bustamante:2020,Li:2022}.
Typically, one end of the molecular system is fixed in space and the other end is pulled away with a constant velocity.
The characteristic forces needed to induce the relevant conformational transformations allow to extract important informations about the  energy landscape of the system\cite{Hyeon:2005,Bustamante:2005}.
Additionally, the transition rates can be computed from the rupture forces employing models of diffusive barrier crossing for the 
kinetics\cite{Hummer:2003,Dudko:2006,Dudko:2008}.

As in many fields of soft-matter physics, computer simulations are extremely helpful in the interpretation of experimental data and allow to gain inside into the molecular dynamics with atomistic resolution\cite{Lee:2009,Luitz:2015}.
Force probe molecular dynamics (FPMD) simulations, also called steered molecular dynamics simulations, are routinely applied not only in order to resolve the conformational transitions in biomolecules but also to test theoretical concepts of statistical 
mechanics\cite{Isralewitz:2001,Park:2004,Sotomayor:2007,Chen:2011}.
However, apart from a recent example\cite{Rico:2013}, usually the time scales of simulations and experiments differ by up to 5 orders of magnitude\cite{Franz:2020}.
The most important reason for the high computational cost of FPMD simulations is the time-consuming calculation of the solvents time evolution, in particular because of the necessity to use large simulation boxes.
Consequently, a number of coarse graining (CG) techniques, such as using implicit solvent models\cite{Bureau:2015}, adaptive 
schemes\cite{Ozer:2012}, structural CG as in Go-models\cite{Li:2009} or polymer models\cite{Zhuravlev:2016} have been applied successfully
in this type of simulations.
Also Markov state models can be used to extrapolate the dynamics on short time scales to the regime relevant for conformational transitions, albeit with the assumption of Markovian kinetics\cite{Ghosh:2017,G89}.
The more common methods applying CG models work with simplified interaction potentials that reduce the number of degrees of freedom of the system.
However, if the focus is on the investigation of the details of molecular conformational changes, an atomistic description of the molecular construct under consideration is advantageous.

In a previous publication\cite{G100}, denoted as I in the following, we performed FPMD simulations in the framework of the adaptive resolution scheme (AdResS) developed by Praprotnik et al.\cite{Praprotnik:2006,Krekeler:2018}.
In these simulations, the system is decomposed into regions of different resolution.
The molecular complex of interest and some layers of solvent molecules define the region where an all-atom (AA) force field is applied.
This so-called AA region is separated from the CG region by a hybrid region allowing for particle exchange.
The CG potentials in the CG region are calculated using the iterative Boltzmann inversion (IBI) method\cite{Reith:2003}.
We showed that the details of the forced unfolding and refolding of a model system, a \cal dimer, are in excellent agreement with the results of fully atomistic MD simulations provided the AA region in the AdResS is chosen large enough.
For too small atomistic regions either parts of the dimer were not treated properly or the impact of the fast dynamics of the CG solvent was responsible for deviations from the results of atomistic simulations.
The gain in computational efficiency is mainly due to the reduced number of particles that have to be treated atomistically.
Of course, the particular choice of the CG potential is expected to have an impact on the results of AdResS simulations and there is no general recipe for its construction.
The simplest choice is provided by having no interactions at all, i.e. an ideal gas (IG) approximation in the CG region and this approach has been proven to be successful when used in AdResS simulations\cite{Kreis:2015}.
 
In the present paper, we extend our former investigation of the AdResS in a twofold manner.
We use the IG approximation instead of a CG potential in the CG region.
Furthermore, we apply the method not only to the previously investigated \cal dimer, but also to another well-studied system, a \bAla\ dissolved in methanol, which is known to unfold via a stable intermediate and not in a direct two-state manner\cite{G84, G85}.
Apart from the possibility to deal with a protic solvent, this demonstrates that one can also treat systems with more complex unfolding pathways, as is often relevant when biomolecules are considered.

The remainder of the paper is organized as follows. 
In the next section, we present our model systems and some details of the computations.
Section III and section IV contain the presentation of the results obtained and their comparison to atomistic simulations.
The paper closes with some conclusions in section V.
\section*{II. Computational Methods}
\begin{figure}[h!]
\centering
\includegraphics[width=12.0cm]{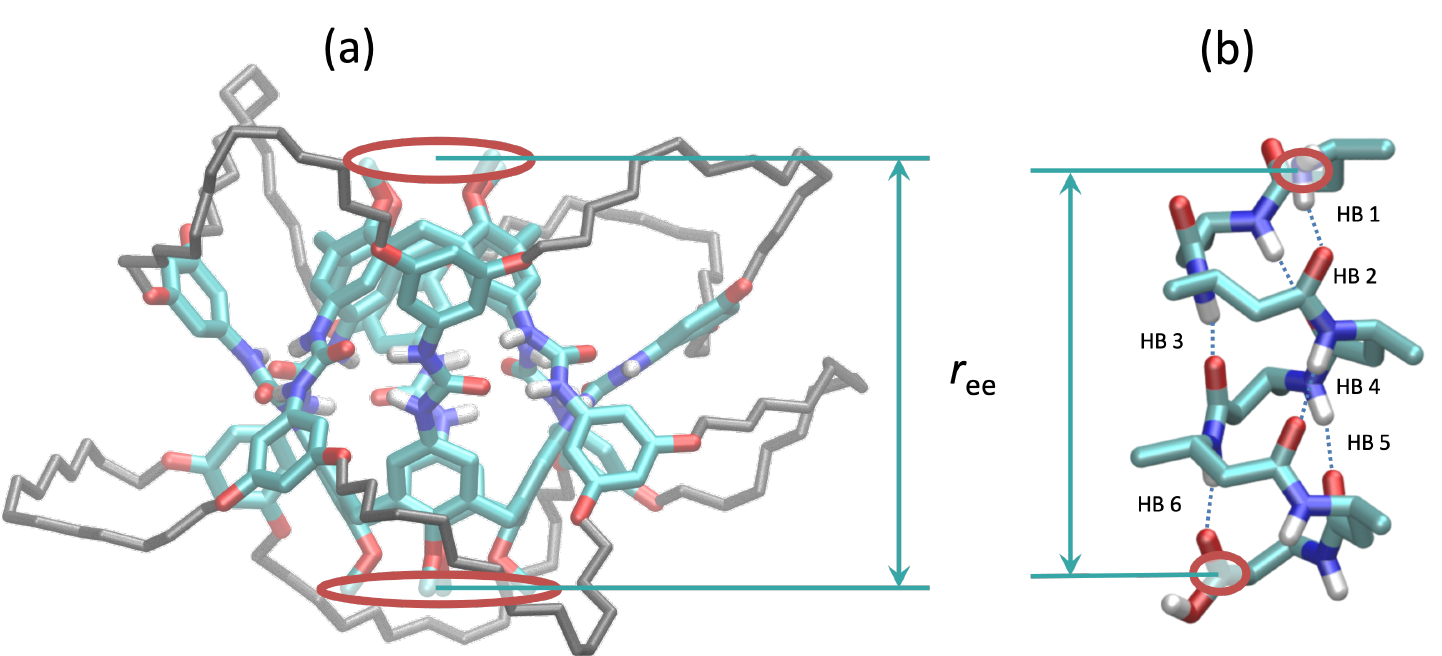}
\caption{{\bf a)} Cartoon of the chemical structure of the \cal catenane dimer.
{\bf b)} Chemical structure of the $\b$-alanine octamer. The dotted lines indicate hydrogen bonds enumerated starting at the N-terminus.
The end-to-end distance $r_{\rm ee}$ is defined as the distance from the reference group to the pulled group.
For the \cal dimer, both groups are defined as the center of mass of the endstanding methyl groups.
In case of the \bAla, the reference group is given by the carbon atom the C-terminus and the pulled group by the nitrogen atom of the N-terminus.
}
\label{Plot1}
\end{figure}

In the following, we describe the details of the simulations performed on the systems investigated, i.e. one \cal dimer dissolved in mesitylene, one \bAla\ in methanol, cf. Fig.\ref{Plot1}, and systems containing only the solvents.
This section partly contains a recapitulation of the corresponding information about our procedure given in paper I.
Throughout the paper, we will use the abbreviations 'Calix' for the \cal dimer in mesitylene and 'bAla' for \bAla\ in methanol.

In equilibrium, the \cal dimer forms a compact structure, denoted as 'closed state', that is stabilized by 16 hydrogen bonds (H-bonds) formed between the urea groups of one monomer with those of the other one (UU-bonds). 
When pulled, the system undergoes a transition to an 'open state' in which 8 H-bonds are formed between the urea group of one monomer and the ether group of the other (UE-bonds)\cite{G69,G74,G77}.
For our purpose, the molecular extension is best characterized by the end-to-end distance $r_{\rm ee}$, as defined in 
Fig.\ref{Plot1}.
In the closed state of the \cal dimer, this is found to be $r_{\rm ee}\simeq1.4$nm and in the open state one has approximately
$r_{\rm ee}\simeq2.2$nm.
We note that the kinetics can be interpreted in terms of a two-state model undergoing reversible transitions, albeit there are some subtleties which, however, are not important for our present discussion\cite{G98}.

As mentioned above, the \bAla\ belongs to the class of $\b$-peptides that have been studied intensively in equilibrium\cite{Cheng:2001} and in FPMD simulations\cite{G84,G85}.
It was found that the pathway of mechanical unfolding in methanol proceeds via a metastable intermediate and the kinetics is best described in terms of three states.
The equilibrium helical structure is stabilized by 6 H-bonds (with $r_{\rm ee}\simeq 1.2$nm) and in the first part of the unfolding transition the outermost bonds are broken and the inner ones only in a second step.
In the stretched conformation, one finds $r_{\rm ee}\simeq 3.5$nm.
This peptide serves as a model system for a more complex transition from a helix to a stretched configuration and we will use it in order to test whether or not the AdResS simulations can be applied successfully also in such a situation.
\subsubsection*{Fully atomistic simulations}
All atomistic simulations (all atom simulations, AA simulations) were performed using the GROMACS 2018.4 program package\cite{Abraham:2015,Abraham:2018} and a stochastic dynamics integrator \cite{Goga:2012} at all temperatures with a friction constant of 0.1 ps.
All bonds were constrained to their equilibrium bond length using the LINCS algorithm\cite{Hess:1997} allowing for a time step of 2 fs.
Short range non-bonding interactions were computed using a cut-off of 1.2 nm.
The long-range Coulomb interactions were treated using the reaction-field method with relative dielectric constants as given in 
Table \ref{MD_param_pure}.
For the van der Waals interactions, we applied a dispersion correction\cite{Allen:1987}. 
The neighbor list was updated after 25 simulation steps.
We used Cartesian periodic boundary conditions in all simulations.

For the neat solvent systems, we performed an energy minimization starting with a (7.5 nm)$^3$ cubic simulation box, followed by an equilibration run in the canonical ensemble for 200 ps.
Then the system was coupled to a Berendsen barostat \cite{Berendsen:1984} with a time constant of 0.5 ps.
All production runs were performed in the canonical ensemble using the box size determined from this equilibration in the NPT-ensemble for a pressure of 1 bar.
In Table \ref{MD_param_pure}, we give the parameters used in the AA simulations of the systems investigated.
\begin{centering}
\begin{table}[h!]
\caption{Parameters used in the AA simulations of the pure solvents}
\begin{tabular}{l|c|c}
\hline
 pure solvent & mesitylene & methanol\\
\hline
force field & OPLS-AA\cite{Hess:2008,opls2,opls1} & MeOH model B3\cite{Walser:2000}\\
\# molecules & 1839 & 6927\\
T / K & 298 & 240\\
isothermal compressibility /bar$^{-1}$ & $8.26\cdot10^{-5}$ & $1.248\cdot10^{-4}$\\
$\epsilon_{\rm rel}$ & 2.4 & 32.7\\
box length / nm & 7.49 & 7.57\\
\end{tabular}\label{MD_param_pure}
\end{table}
\end{centering}

For the \cal dimer, the OPLS-AA force field was used as in our earlier investigations\cite{G100, G77, G88}. 
For the \bAla\ we used the GROMOS 53a6 force field\cite{Oostenbrink:2004} that has been shown to be applicable also in FPMD 
simulations\cite{G84, G85}. 
In the mixtures, the box lengths determined via the NPT equilibration are the same as for the pure solvents, only the number of solvent molecules are changed.
In the Calix system, we have 1780 mesitylene molecules and for the bAla mixture, there are 6919 methanol molecules in the simulation box.
\subsubsection*{Adaptive resolution simulations}
The general setup for the AdResS simulations can briefly be summarized as follows.
Inside a spherical AA region with radius $r_{\rm AA}$ the system is treated atomistically. The CG region outside is separated from the AA region by a hybrid region with a slab thickness $s_{\rm Hy}$.
In this region the forces are interpolated between the AA and the CG part allowing for particle exchange between the regions of different resolution.
For more details see, e.g. paper I\cite{G100}.

For the simulations using the AdResS, in a first step, the CG potentials have to be determined.
The potential for the interaction between mesitylene molecules has been calculated in I using the IBI method.
The same procedure was applied to compute the CG potential for methanol.
As in I, the AdResS simulations were performed using the VOTCA program package\cite{Ruhle:2009, Mashayak:2015} (VOTCA 1.3.).
While for the determination of the CG potential for mesitylene 325 iterations were needed in order to obtain converged potentials, in case of methanol the procedure converged already after 230 iterations.
We mention that the speedup in the dynamics as measured by the increase of the diffusion contant was found to be given by a factor of 
$D_{\rm CG}/D_{\rm AA}\simeq6.9$ for mesitylene and $D_{\rm CG}/D_{\rm AA}\simeq13.4$ for methanol.
The thermodynamic force was calculated for different sizes $r_{\rm AA}$ and a constant slab thickness of the hybrid region of 
$s_{\rm Hy}=1.2$.
The AdResS simulations on the bAlA system using the CG potentials determined this way were performed in an identical manner as in I.

The procedure used to compute the thermodynamic force is independent of the particular choice of the CG potential and only depends on the difference in density\cite{Fritsch:2012}.
Therefore, in case of the IG approximation, we proceeded in the same way as in case of the CG potentials determined with the IBI method.
Some additional details of the calculations are presented in Appendix A.

In I, we have discussed that the AdResS simulations employing the IBI method for the calculation of the CG potential results in a speedup compared to AA simulations.
We estimated that roughly half of the solvent molecules have to be treated atomistically, if $r_{\rm AA}=1.6$nm was chosen.
Using the IG approximation instead of the CG potential determined using IBI does not reduce the computational cost of the simulation itself.
However, one does not have to compute any CG potential at all.
\subsubsection*{Force probe simulations}
The FPMD simulations have always been performed according to the same protocol independent of the choice of the general simulation setup.
One end of the molecule was fixed in space (reference group) and the other end was pulled apart (pulled group) with a constant velocity.
For this purpose, a harmonic potential has been applied to the pulled group and the force measured at the position of the spring is given by
\be\label{F.def}
F = K(v\cdot t - z(t)) = \mu\cdot t - K\cdot z(t) .
\ee
Here, $K$ is the spring constant, $v$ the pulling velocity and $\mu=K\cdot v$ is the loading rate.
Furthermore, $z(t)$ denotes the deviation of the position of the pulled group from its initial value. 
The reference group and the pulled group for the \cal dimer and the \bAla\ are defined in the caption to Fig.\ref{Plot1}.
In each case, the system was pulled until a maximum extension $x_{\rm max}$ was reached, $4$nm in case of the \cal dimer and $3.5$nm for the \bAla.
In the former case, the simulations were performed in two modes, the pull mode just described and additionally a relax mode.
This means, that after reaching $x_{\rm max}$, the pulling direction was inverted with all other parameters remaining unaltered.

\begin{figure}[h!]
\centering
\vspace{-0.25cm}
\includegraphics[width=8.0cm]{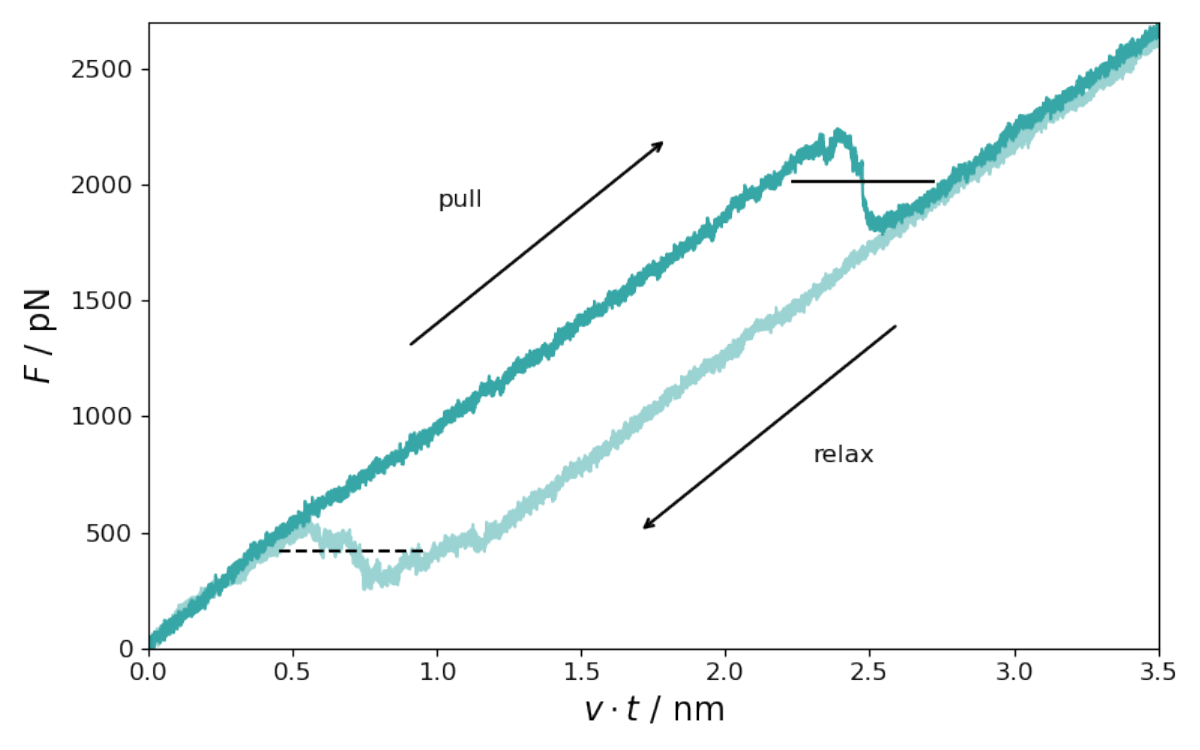}
\vspace{-0.5cm}
\caption{Example of an FEC monitored via an AA simulation for the Calix system.
The pulling velocity was $v=1$m/s and the spring constant was $K=1$N/m.
The black horizontal line indicates the rupture force and the dashed line the rejoin force.
}
\vspace{-0.25cm}
\label{C_FEC_AA}
\end{figure}

A typical force versus extension curve (FEC) obtained using this protocol for an AA simulation of the Calix system is presented in 
Fig.\ref{C_FEC_AA}.
The force increases (linearly) until the system is stretched so strong that a conformational transition, in most cases accomponied by the opening of one or more H-bonds, occurs. 
Immediately after the transition, characterized by the rupture force, the force drops because the molecular construct elongates and therefore the spring of the pulling device relaxes.
This drop is followed by a further increase of the force for still larger extensions.
In the relax mode the force decreases and at some point the transition to the equilibrium structure takes place defining the rejoin force.
The forward and backward transitions would only take place at the same value of the extension in the quasistatic limit. 
With the velocities encountered in FPMD simulations, one usually observes a hysteresis between the FECs in the two different modes.

\section*{III. Calix[4]arene dimer in mesitylene - Calix}
It was shown in paper I, that the AdResS methodology can be applied to FPMD simulations also in the case of fast pulling provided the atomistic region is chosen large enough.
Before we discuss the results obtained using the IG approximation in the AdResS simulations, we briefly recapitulate the main findings of I for convenience of the reader.

In equilibrium, the results of the AdResS simulations for $r_{\rm AA}\geq0.8$nm coincide with the AA simulations in the sense that the distributions of the end-to-end distances $r_{\rm ee}$ and also the number of UU-bonds stabilizing the closed structure of the dimer are well represented.
The FPMD simulations were performed as described above and were analyzed in detail regarding different kinetic and structural parameters.
Because molecular conformational transitions are of a stochastic nature, 300 simulations were performed for each set of parameters, allowing for a meaningful statistical analysis.
For a value of $r_{\rm AA}=1.6$nm, the AdResS simulations coincided with the AA simulations also for very fast pulling.
This means that the averaged FECs and the distributions of rupture forces and rejoin forces agree within statistical uncertainties.
For smaller AA regions, the enhanced mobility of the solvent molecules in the hybrid region and in the CG region gives rise to a softening of the molecular energy landscape manifesting itself in smaller rupture forces and larger rejoin forces.
The number of H-bonds, both, the UU-bonds in the closed state and the UE-bonds in the open state are well represented by the AdResS simulations with $r_{\rm AA}=1.6$nm.
This holds in the pull mode and also in the relax mode.

\begin{figure}[h!]
\centering
\vspace{-0.25cm}
\includegraphics[width=14.0cm]{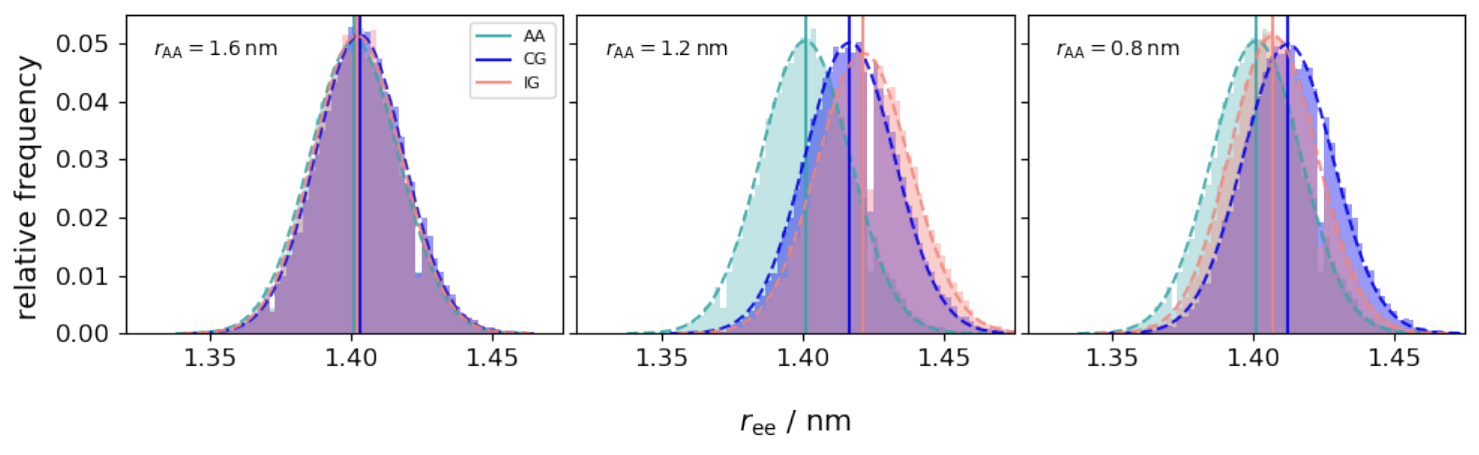}
\vspace{-0.5cm}
\caption{Distribution of end-to-end distances $r_{\rm ee}$ as determined from equilibrium simulations for various sizes of the atomistic region, $r_{\rm AA}$, in AdResS simulations.
}
\vspace{-0.25cm}
\label{C_Ree}
\end{figure}

In the following, we compare these results with those obtained in the present study for an IG approximation in the CG region.
In Fig.\ref{C_Ree}, we present the distributions of the end-to-end distance $r_{\rm ee}$ of the \cal dimer as obtained from equilibrium simulations for the various setups.
It is evident from this plot that the results for the different CG potentials are very similar to each other for all values of $r_{\rm AA}$.
Additionally, for both CG strategies one obtains results that cannot be distinguished from the AA simulation results for $r_{\rm AA}=1.6$nm.

\begin{figure}[h!]
\centering
\vspace{-0.25cm}
\includegraphics[width=12.0cm]{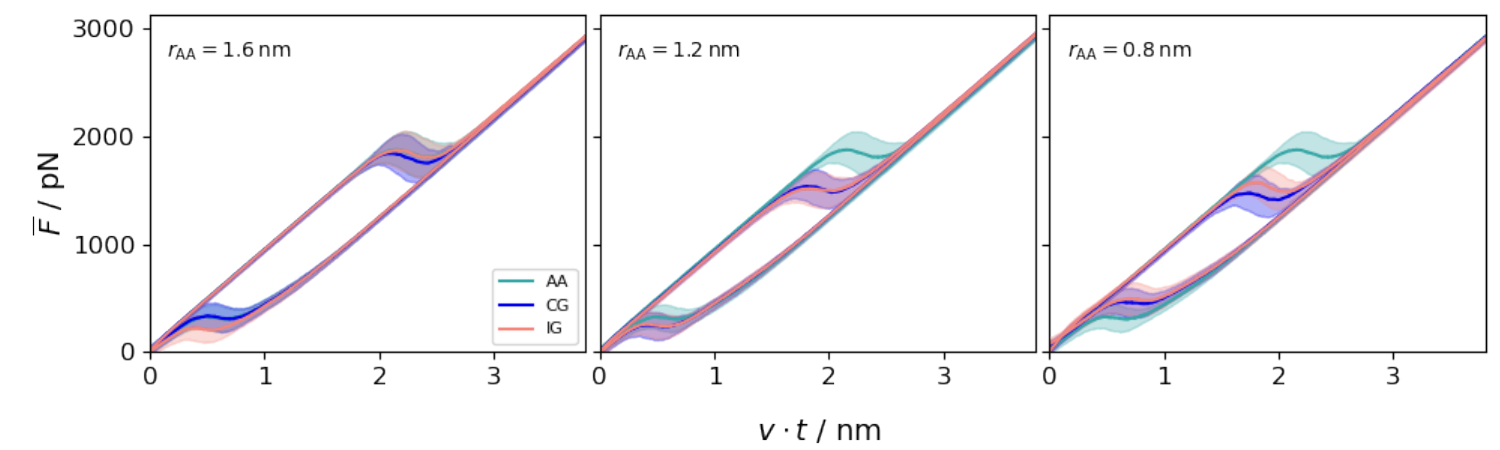}
\vspace{-0.5cm}
\caption{Averaged FECs for both, the pull mode and the relax mode, as a function of the extension for various sizes of the atomistic region, $r_{\rm AA}$.
Here, the pulling velocity was $v=1$m/s and the spring constant was $K=1$N/m and the average was performed using 300 individual FECs.
The shaded areas indicate the spread of the individual FECs (second moment of the distributions).
}
\vspace{-0.25cm}
\label{C_FEC}
\end{figure}

The FPMD simulations have been performed for the same parameters (pulling velocities and spring constants) as in I and we present examples of averaged FECs (also denoted as dynamic strength) is Fig.\ref{C_FEC}.
In contrast to the equilibrium simulations, one observes rather strong discrepancies between the different setups for small atomistic regions, particularly in the regime of the closed-to-open transition of the \cal dimer.
The slopes of the FECs are independent of the CG method and in all cases coincide with those obtained from AA simulations.
This means, that the molecular stiffness remains unaltered, independent of the approximation used for the CG potential for the solvent.
We mention, that this robustness of the molecular spring constant has also been observed in FPMD simulations employing a hybrid scheme with virtual sites for the interaction between the \cal dimer treated atomistically and the mesitylene solvent in a CG representation\cite{G88}.

\begin{figure}[h!]
\centering
\vspace{-0.5cm}
\includegraphics[width=10.0cm]{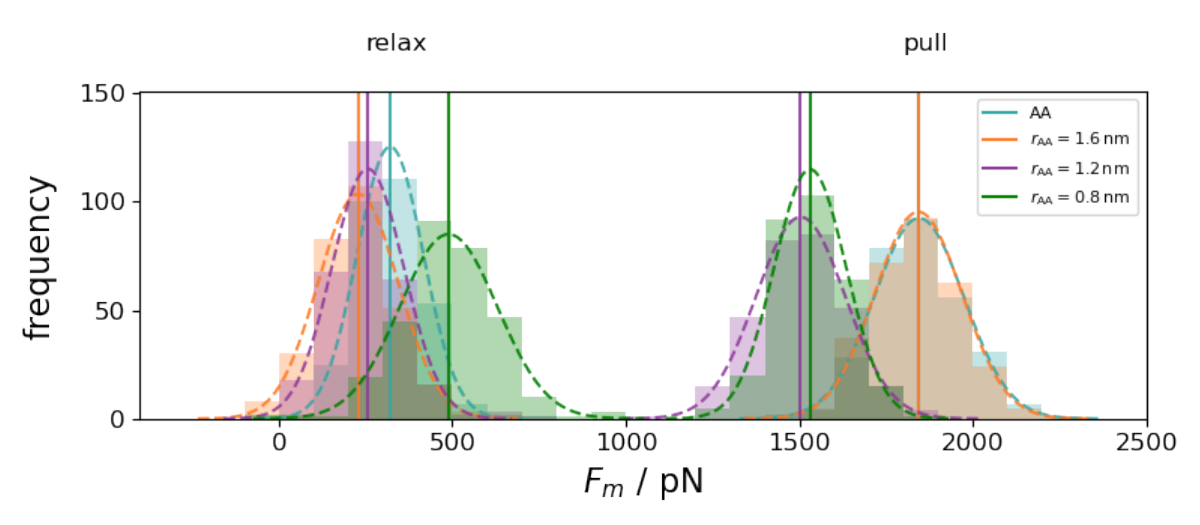}
\vspace{-0.5cm}
\caption{Distributions of rupture forces (pull mode) and rejoin forces (relax mode) as defined in Fig.\ref{C_FEC_AA} using the IG approximation in the treatment of the solvent in the CG region for $v=1$m/s and $K=1$N/m.
Dashed lines are fits to a Gaussian.
}
\vspace{-0.25cm}
\label{C_Fdist}
\end{figure}

From the individual FECs, we extracted the distributions for the rupture forces and the rejoin forces and present the results in 
Fig.\ref{C_Fdist}.
Both, the width and the mean of the distributions obtained for $r_{\rm AA}=1.6$nm coincide very well with those of the AA simulation.
In the pull mode, the deviations for the results for $r_{\rm AA}=1.2$nm are somewhat larger than in the relax mode.

From the results presented here, we conclude that the IG approximation for the solvent in the CG region is very well suited to speed up molecular simulations of a system exhibiting reversible bond breaking even in the case of fast pulling.
It is only required that the AA region is chosen large enough.
If this is fulfilled, the simplest possible choice for coarse graining is sufficient.
\section*{IV. \bAla\ in methanol - bAla}
As discussed above, the \cal dimer can be viewed as a simple two-state model for reversible bond formation.
However, in most applications of force spectroscopy, one has to deal with systems exhibiting a more complex (un)folding pathway.
Additionally, mesitylene is of an aprotic nature and therefore the solvent does not interfere with the formation of H-bonds in the dimer.
In the present study, we use the bAla system in order to test the applicability of the AdResS method in a situation where the mechanical unfolding proceeds via a metastable intermediate and additionally the solvent is able to form competing H-bonds.

\begin{figure}[h!]
\centering
\vspace{-0.25cm}
\includegraphics[width=10.0cm]{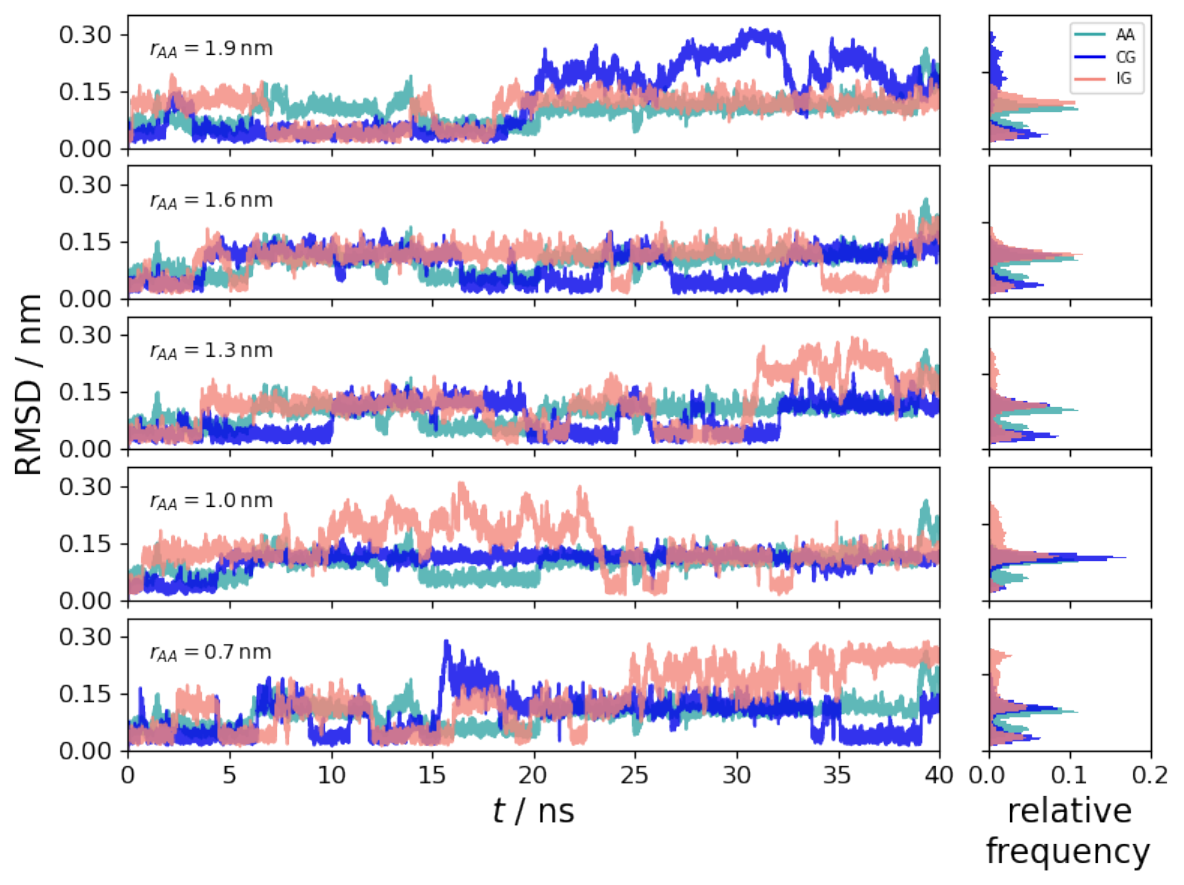}
\vspace{-0.25cm}
\caption{RMSD for an equilibrium simulation for different sizes of the AA region as indicated.
The right panels show the distributions of the RMSD.
}
\vspace{-0.25cm}
\label{A_RMSD}
\end{figure}

The $3_{14}$-helix formed by the \bAla\ is stabilized by 6 H-bonds. 
At $T=240$K there are some fluctuations in the number of H-bonds that are closed and therefore the helical character of the peptide structure also changes to some extent in the course of time.
A common measure for this type of fluctuations is provided by the root mean square deviation (RMSD) of the actual time-dependent structure from to a given reference structure.
In Fig.\ref{A_RMSD}, we present the RMSD as a function of the simulation time for the different simulation setups, i.e., AA simulations,
AdResS simulations using the CG potential, and those applying the IG approximation.
The reference structure in all simulations is the perfect helix at the beginning of the trajectory.
In the right panels, the distributions of the RMSD are shown.
It is apparent from this figure, that the helix is quite stable with a second maximum of the RSMD at about 0.15nm corresponding to some structures in which the H-bond at the N-terminus (H-bond 1), cf. Fig.\ref{Plot1}, is opened.
Furthermore, the AdResS simulations give a reasonable representation of the main features observed in the AA simulations, independent of the choice of $r_{\rm AA}$.
We conclude that the AdResS simulations yield reliable results in equilibrium even for rather small AA regions.

\begin{figure}[h!]
\centering
\vspace{-0.25cm}
\includegraphics[width=9.0cm]{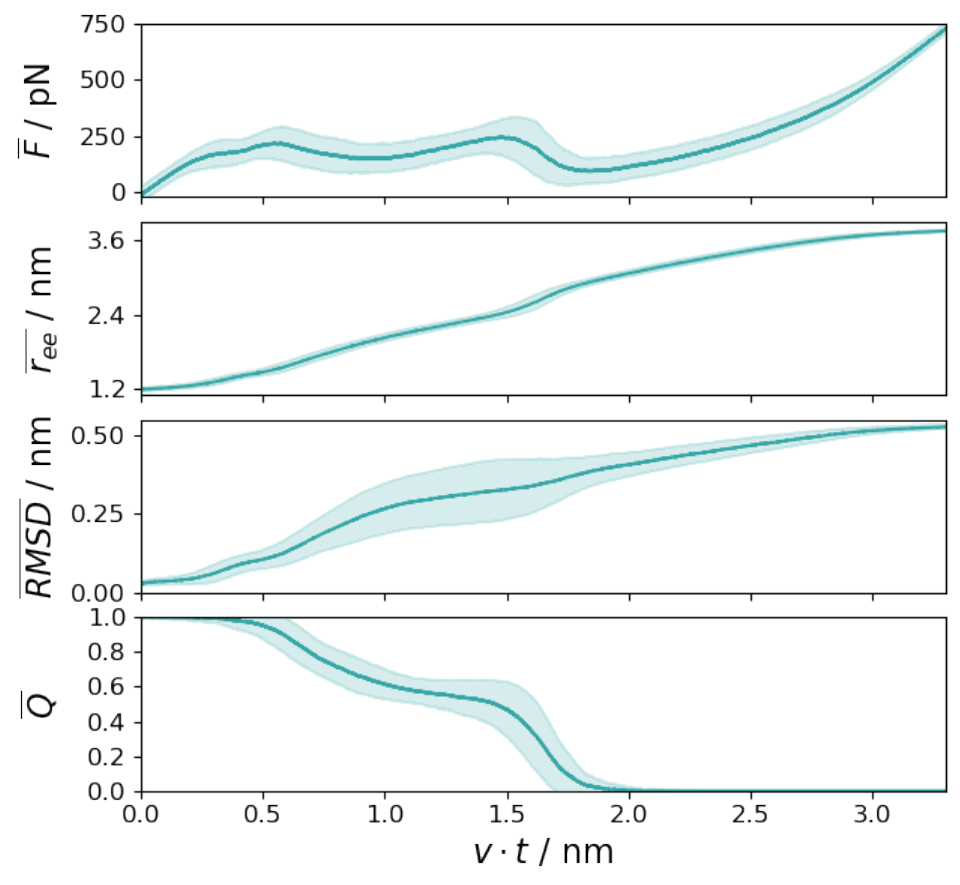}
\vspace{-0.25cm}
\caption{Mean rupture force, mean end-to-end distance, mean RMSD, and mean fraction of native contacts as a function of the extension.
The shaded areas are determined by the second moments of the respective distributions.
}
\vspace{-0.25cm}
\label{A_OP}
\end{figure}

\begin{figure}[h!]
\centering
\vspace{-0.25cm}
\includegraphics[width=11.0cm]{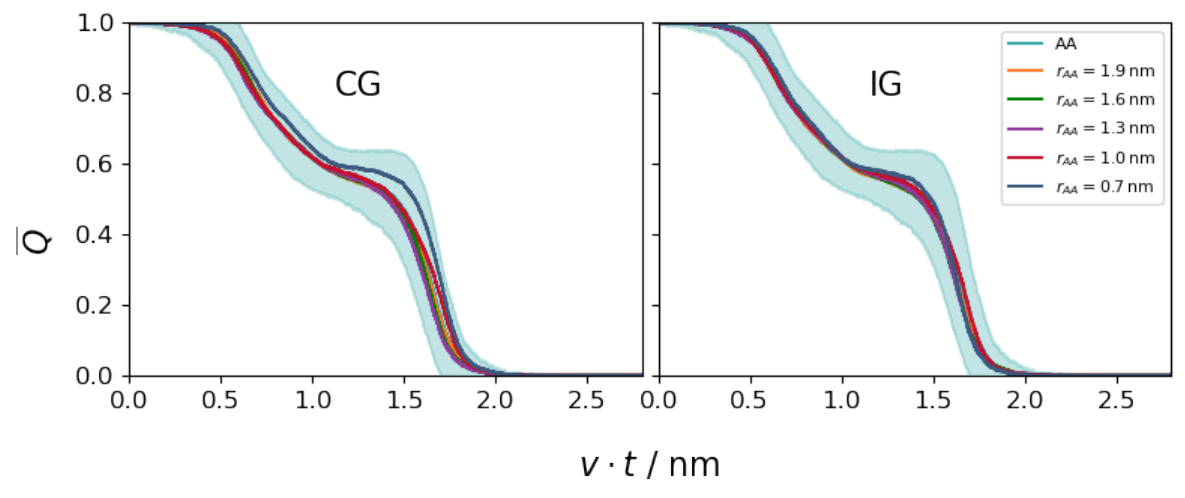}
\vspace{-0.25cm}
\caption{Mean fraction of native contacts, $\overline{Q}$, as a function of the extension for the AdResS simulations as indicated.
The shaded area are represents the width of the distribution for the AA simulation.
}
\vspace{-0.25cm}
\label{A_Qext}
\end{figure}

Next, we will discuss the results of FPMD simulations.
Throughout this Section, we discuss the results obtained from 1000 FPMD simulations using $v=1$m/s and $K=1$N/m corresponding to a very large loading rate of $\mu=1$N/s.
In the case of bAla, the situation is somewhat more involved than for the Calix system, where it is sufficient to consider the end-to-end distance, $r_{\rm ee}$, in order to characterize the reversible opening transitions of the dimer.
It is known from earlier atomistic constant-force MD simulations that the end-to-end distance is not a well-defined order parameter for the bAla system\cite{Knoch:2017}.
In Fig.\ref{A_OP} we compare different variables as a function of the extension for AA simulations.
The overall structure of the quantities shown in Fig.\ref{A_OP} is determined by the fact that the unfolding pathway of the \bAla\ in FPMD simulations proceeds via a metastable intermediate\cite{G84,G85}.
The mean FEC roughly exhibits two force maxima indicating the two unfolding transitions.
However, the rips in the FECs are less pronounced than in case of the Calix system, cf. Fig.\ref{C_FEC}.
In the mean values of the end-to-end distance and the RMSD, the two unfolding events are represented only weakly.
On the other hand, the mean fraction of native contacts\cite{G84,Best:2013}, $\overline{Q}$, clearly exhibits a plateau at extensions around 1.2nm.
It therefore might be used for a characterization of the unfolding pathway even for such small systems as the \bAla.
In Fig.\ref{A_Qext}, we present $\overline{Q}$ for the various AdResS simulations.
It can be seen that for all sizes of the AA region except the smallest one, the results of both types of AdResS simulations, CG potential and IG approximation, give very good representations of the overall unfolding pathway in the sense that the intermediate is observed in both cases.

\begin{figure}[h!]
\centering
\vspace{-0.25cm}
\includegraphics[width=11.0cm]{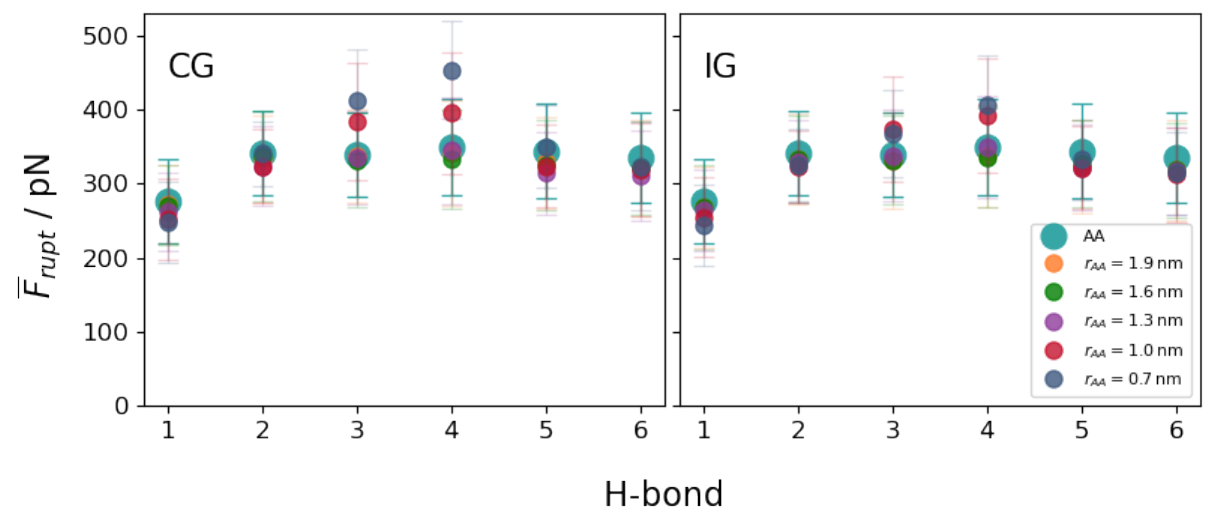}
\vspace{-0.25cm}
\caption{Mean rupture forces for all 6 H-bonds as defined in Fig.\ref{Plot1}.
The error bars are determined by the second moments of the respective distributions.
}
\vspace{-0.25cm}
\label{A_FHB}
\end{figure}

In a next step, we extracted the rupture forces for each H-bond from the FECs.
For this purpose, we defined a H-bond as intact for a donor-acceptor distance of less than 0.35nm and the angle formed by the ON- and the OH-bonds of less than 30$^o$\cite{G74}. 
In Fig.\ref{A_FHB}, we show the averaged rupture forces for each individual H-bond for the three different simulation setups.
These forces were determined from the last opening of a given H-bond during the pulling trajectory.
One can see that only for the smallest value of $r_{\rm AA}$ there are deviations of the values obtained from the AdResS simulations when compared to the AA simulation results.
We note that results of similar quality are obtained for other pulling parameters.
In Appendix B, we present the distributions of the rupture forces for all simulation setups and various values of $r_{\rm AA}$.

\begin{figure}[h!]
\centering
\vspace{-0.25cm}
\includegraphics[width=15.0cm]{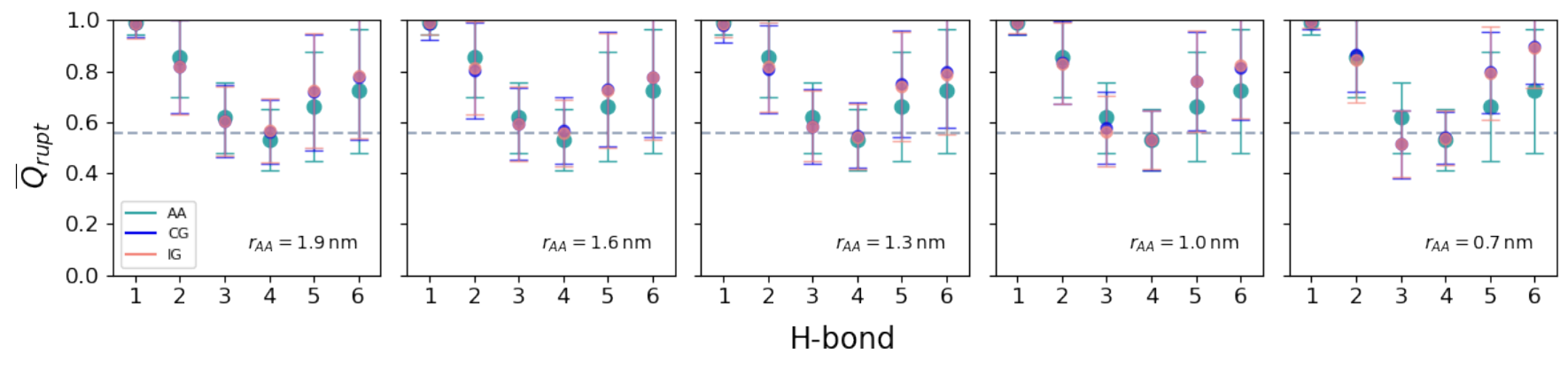}
\vspace{-0.25cm}
\caption{Mean value of the fraction of native contacts at rupture for all 6 H-bonds as defined in Fig.\ref{Plot1}.
The error bars are determined by the second moments of the respective distributions.
The dashed horizontal lines are a guide to the eye.
}
\vspace{-0.25cm}
\label{A_QHB}
\end{figure}

\begin{figure}[h!]
\centering
\vspace{-0.25cm}
\includegraphics[width=15.0cm]{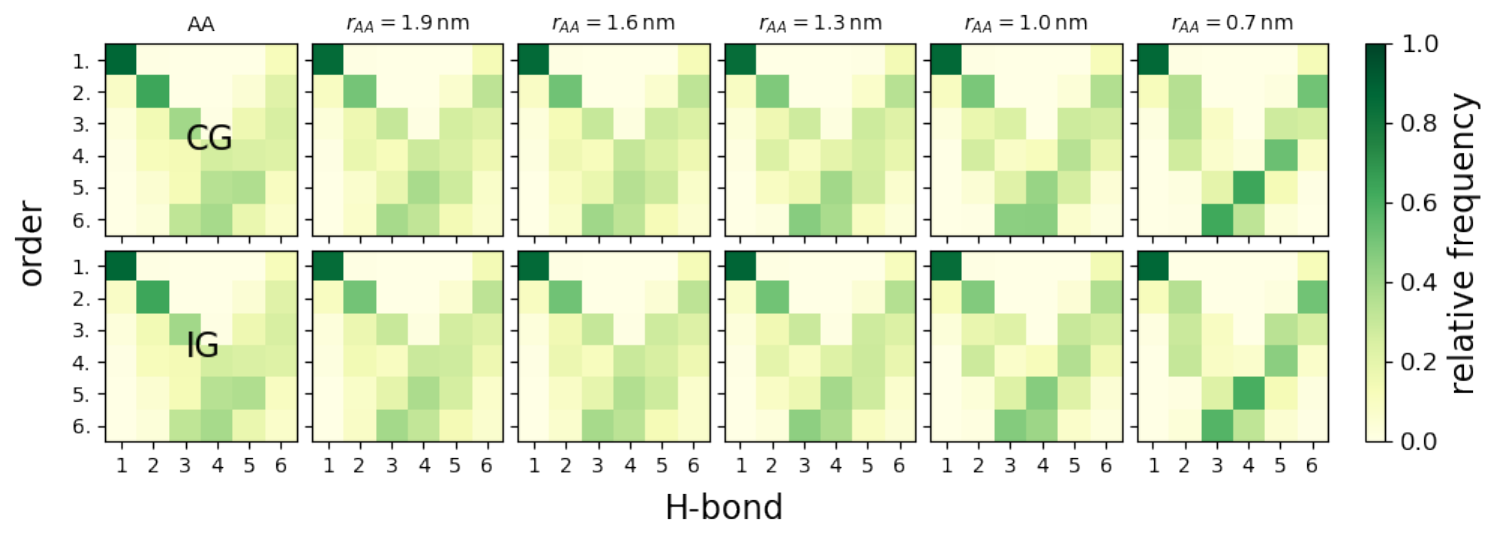}
\vspace{-0.25cm}
\caption{Relative order of the H-bond opening for 1000 FPMD trajectories. Upper panels: CG potential; Lower panels: IG approximation.
}
\vspace{-0.25cm}
\label{A_HBorder}
\end{figure}

As mentioned above in the context of Figs.\ref{A_OP},\ref{A_Qext}, the fraction of native contacts can be used to characterize the unfolding pathway.
In Appendix B, by considering the averaged FECs as a function $\overline{Q}$, we additionally show that the unfolding pathway does not depend on the pulling velocity.
In Fig.\ref{A_QHB}, we present the mean value of the fraction of native contacts at that value of the extension at which a given H-bond breaks,
$\overline{Q}_{\rm rupt.}=\overline{Q}(r_{\rm ee}({\rm rupt.}))$.
It is obvious, that the H-bond located at the N-terminus is the first to open.
This is consistent with the observation that also in equilibrium this H-bond is the weakest.
More important is the finding that the intermediate state in all cases is stabilized by the innermost H-bonds (nr. 3 and nr.4).
This holds independent of the approximations used in the simulations and can be taken as an indication for the fact that the unfolding takes place along the same path.
We can further investigate to which extent also the correct order of the opening of the H-bonds is captured by the AdResS simulations.
For this purpose, we plot the relative frequency of the order of the opening of the different H-bonds in Fig.\ref{A_HBorder}.
As expected, H-bond 1 is the first to open in all simulations irrespective of the approximation applied.
Furthermore, for $r_{\rm AA}\geq1.3$nm the AdResS simulations give excellent results, only for smaller AA regions there are some deviations in the order of the opening transitions.
Also the finding that the innermost H-bonds (3 and 4) are the more stable ones is recovered by the order of the transitions since they open very late in most of the simulations.
The most important result of these simulations is that the AdResS method approximates the results of the AA simulations also in this case of an unfolding pathway exhibiting a metastable intermediate in a protic solvent.
\section*{IV. Conclusions}
It is well known that AdResS simulations give excellent results when compared to AA simulations for various systems in equilibrium and we have also observed this agreement between AdResS and AA simulations, both, in our previous study and in the present paper, even for very small AA regions.

In the present study, we applied the AdResS methodology to perform FPMD simulations for different parameters like pulling velocities.
For one system, a \cal dimer dissolved in mesitylene, in our earlier work we had shown the applicability of the method in the particular non-equilibrium situation encountered in systems under the influence of a time-varying external force.
This dimer shows reversible one-step folding/unfolding and therefore provides a very simple example of a conformational transition.

In the present paper, we investigated two important aspects of the application of the AdResS method to FPMD simulations.
In addition to the \cal dimer, we studied a second model system, a \bAla\ in methanol solvent, which is known to undergo a two-step unfolding transition under the influence of an external force.
By carefully analyzing the statistics of the fraction of native contacts, the characteristic rupture forces, and the order of the opening of the H-bond stabilizing the helix, we show that the unfolding pathway remains unaltered provided the AA region is chosen large enough.
This is encouraging, in particular because methanol is a protic solvent and therefore electrostatic interactions are expected to be more important than in the case of mesitylene.

The second goal of the present study was to investigate the impact of the methodology used to compute the CG potential.
The choice and the calculation of the CG potential is not unique and the comparison to atomistic simulations is the ultimate test for the quality of the results obtained using a given CG potential.
We demonstrate that even the simplest possible choice for the coarse graining procedure, namely the IG approximation, gives reliable results when used in FPMD simulations.
The only two prerequisites for the method to work are the careful calculation of the thermodynamic force in order to obtain a density match between the regions of different resolution and the choice of an AA region of sufficient size.
Of course, the latter is determined by the system at hand and has to be adjusted carefully.
Apparently, it is sufficient to treat the interactions between the solute and the first few shells of neighboring solvent molecules atomistically in order to compensate the effect of the dynamical speedup of the dynamics of the solvent in the CG region.

In conclusion, we have unequivocally shown that the AdResS employing the IG approximation instead of a CG potential can be applied in FPMD simulations, also for fast pulling.
There are a few directions for future investigations.
For instance, the implementation of non-spherical geometries and adaptive sizes of the AA region might speed up the simulations for larger systems, where the differences in the end-to-end distances of the folded structure and the stretched conformation are large.
\section*{Acknowledgement}
Financial support by the DFG via TRR 146 (project number B3) is gratefully acknowledged.
The authors gratefully acknowledge the computing time granted on the supercomputer Mogon at Johannes Gutenberg University Mainz (hpc.uni-mainz.de).
\newpage
\begin{appendix}
\section*{Appendix A: CG potential for methanol}
\setcounter{equation}{0}
\renewcommand{\theequation}{A.\arabic{equation}}
\setcounter{figure}{0}    
\renewcommand\thefigure{A.\arabic{figure}} 
The effective interaction potential for methanol was computed in the same way as for mesitylene in paper I.
The IBI procedure starts from the potential of mean force for a two body potential, $U(r)=-k_BT\ln(g(r))$.
Here, $g(r)$ is the radial distribution function (RDF) and $k_B$ the Boltzmann constant.
The effective pair potential is computed in an iterative manner:
\be\label{PMF.CG}
U(r)^{CG}_{i+1}=U(r)^{CG}_i+k_BT\ln\left({g(r)_i\over g(r)_{ref}}\right).
\ee
In this expression, $g(r)_{ref}$ denotes the reference RDF of the atomistic liquid, i.e. the RDF of the center of mass of the methanol molecules, and it determines the initial potential, $U(r)^{CG}_0=-k_BT\ln(g(r)_{ref})$.
As in I, we used a cut-off for the potential of $1.2$nm and each simulation run had a duration of 200 ps (the first 20 ps were omitted for equilibration).
$U(r)^{CG}$ was smoothed using cubic splines and we iterated the procedure 230 times without pressure correction.
The simulations were performed using the program package VOTCA\cite{Ruhle:2009, Mashayak:2015} (VOTCA 1.3.)
The results for the RDF are presented in Fig.\ref{A_CGPot}. 
\begin{figure}[h!]
\centering
\vspace{-0.25cm}
\includegraphics[width=7.0cm]{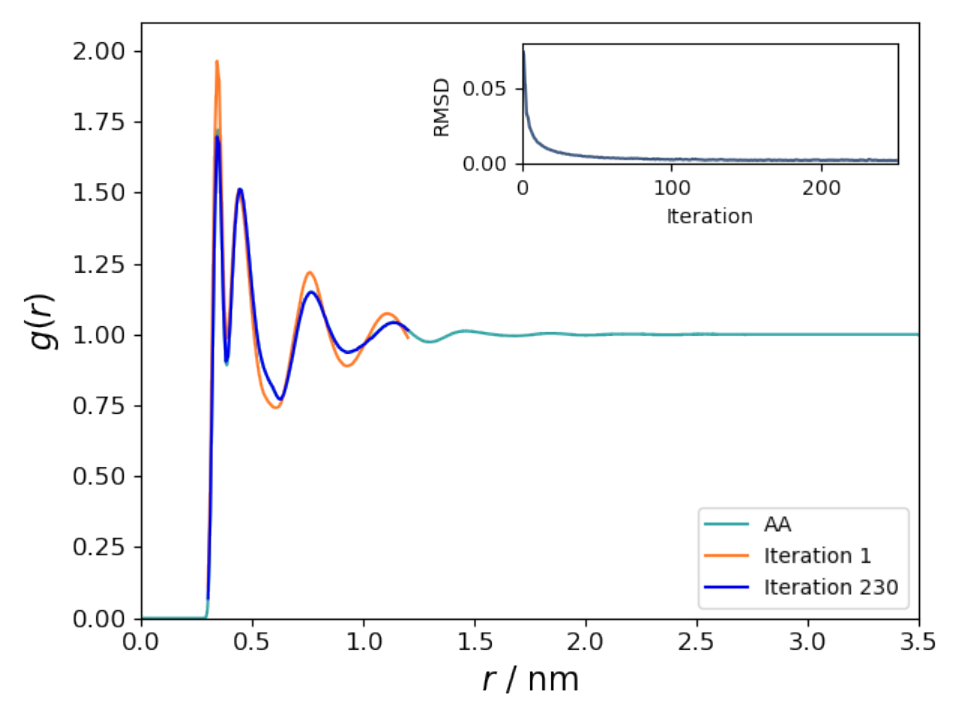}
\hspace{1cm}
\includegraphics[width=7.0cm]{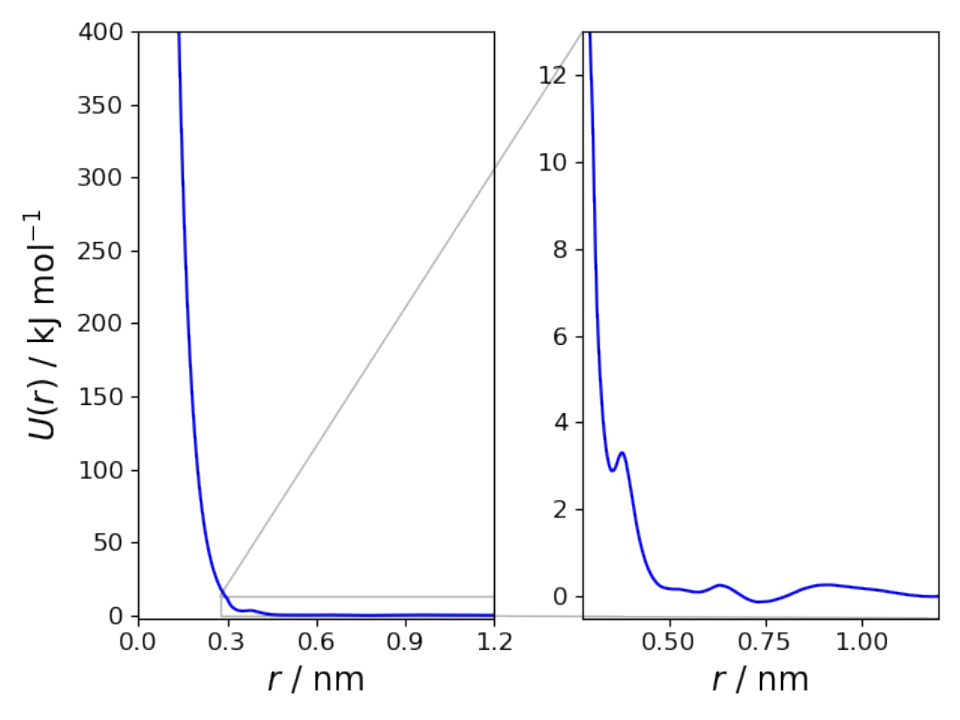}
\vspace{-0.25cm}
\caption{left panel: Evolution of the RDF in the IBI method as indicated. The inset shows the RMSD of $gr)_i$ relative to $g(r)_{ref}$.
Right panels: Resulting CG potential.
}
\vspace{-0.25cm}
\label{A_CGPot}
\end{figure}
It is evident that the resulting CG potential is mainly of a repulsive nature.
\section*{Appendix B: Details of \bAla\ unfolding}
\setcounter{equation}{0}
\renewcommand{\theequation}{B.\arabic{equation}}
\setcounter{figure}{0}    
\renewcommand\thefigure{B.\arabic{figure}} 
In the main text we show the mean rupture forces for each H-bond of the \bAla\ in Fig.\ref{A_FHB}.
These forces were determined from the distributions of the rupture forces computed from analyzing 1000 FECs.
For this reason, we determined the force at which each H-bond opens according to the criteria mentioned in the text.
The resulting distributions for the different types of simulations performed are presented in Fig.\ref{A_Fr_all}.
\begin{figure}[h!]
\centering
\vspace{-0.25cm}
\includegraphics[width=14.0cm]{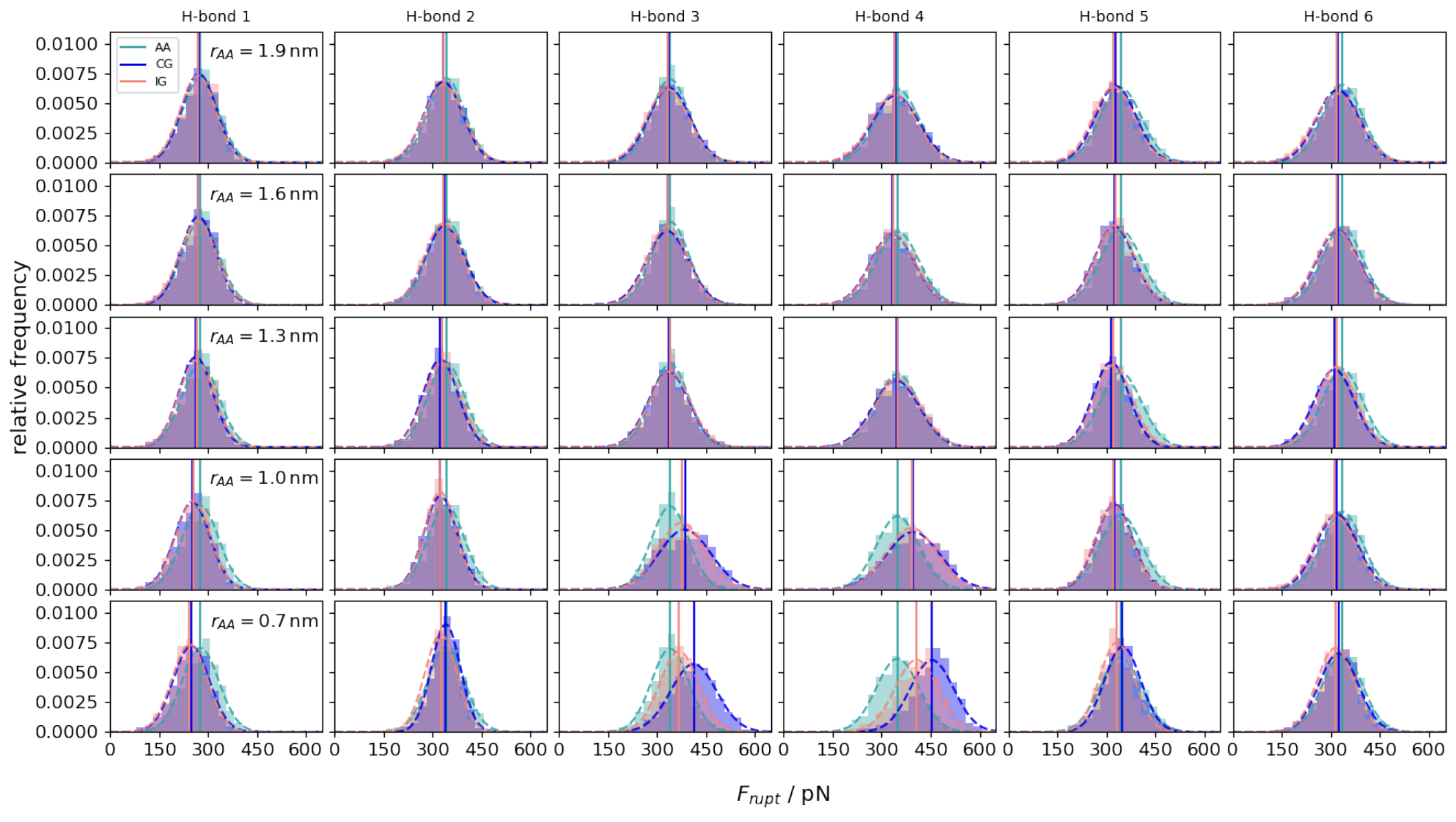}
\vspace{-0.25cm}
\caption{Distributions of rupture forces for the individual H-bonds and various sizes of the AA region.
Broken lines are fits to a Gaussian.
The CG potential was calculated using the IBI method.
}
\vspace{-0.25cm}
\label{A_Fr_all}
\end{figure}
It is evident that not only the mean values but also the shape of the distributions represent the AA simulations very well for a large enough AA region.

\begin{figure}[h!]
\centering
\includegraphics[width=9.0cm]{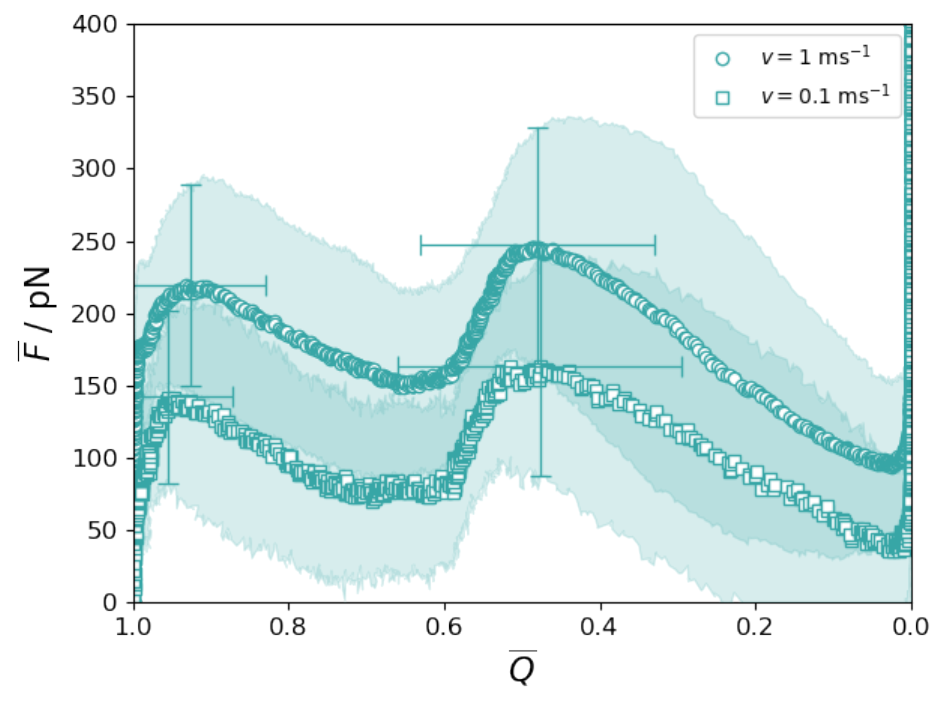}
\vspace{-0.25cm}
\caption{Averaged FEC versus mean fraction of native contacts for AA simulations of two different pulling velocities and $K=1$N/m.
The shaded area is the width of the distribution of FECs (second moment) and the crosses indicate the mean values of the maximum force along with the breadth from the distributions.
}
\vspace{-0.25cm}
\label{A_Fm_Q}
\end{figure}

In the main text, it was found that the fraction of native contacts can serve as a meaningful order parameter for the unfolding of the
\bAla\ in methanol.
This fact can be used in order to show that the unfolding pathway does not depend on the pulling velocity.
For this purpose, we present the averaged FEC  as a function of $\overline{Q}$ in Fig.\ref{A_Fm_Q}.
It is evident from this figure, that the maxima in the averaged FECs are located at very similar values of $\overline{Q}$ and only the value of the force maximum depends on the pulling velocity.
The only effect of varying the velocity is to shift the whole curves while the shape remains unaltered.
\end{appendix}
\newpage
\end{document}